\documentclass[aps,prl,twocolumn,amsmath]{revtex4}
\usepackage[normalem]{ulem} 
\usepackage{color}

\definecolor{g-blue}{rgb}{0.83,0.95,1}
\definecolor{g-yellow}{rgb}{1,1,0.7}
\definecolor{g-green}{rgb}{0.9,1,0.9}
\definecolor{green}{rgb}{0,0.6,0}
\definecolor{cyan}{rgb}{0,0.7,0.7}
\definecolor{black}{rgb}{0,0,0}
\definecolor{grey}{rgb}{0.4 ,0.4 ,0.4 }

\usepackage{bm}
\usepackage{amssymb}
\usepackage{amsfonts}
\usepackage{amsmath}
\usepackage{graphicx}
\usepackage{amsmath,bm}
\def \ed {\end{document}}
\def\Fbox#1{\vskip1ex\hbox to 8.5cm{\hfil\fboxsep0.3cm\fbox{%
		\parbox{8.0cm}{#1}}\hfil}\vskip1ex\noindent}  %%  {TEXT} in BOX

\newcommand{\Eqs}[1]{Eqs.\,(\ref{#1})}%%  requires \eq{label}
\newcommand{\Fig}[1]{Fig.\,\ref{#1}}%%  requires \Fef{label}
\newcommand{\Ref}[1]{Ref.\,\cite{#1}}%%  requires \Fef{label}
\def\be{\begin{equation}}
\def\ee{\end{equation}}
\def\bea{\begin{eqnarray}}
\def\eea{\end{eqnarray}}
\def\bse{\begin{subequations}}
\def\ese{\end{subequations}}

\def\rn{{\rm n}}
\def\rs{{\rm s}}

\let \= \equiv  \let\*\cdot \let\~\widetilde \let\^\widehat \let\-\overline
\let\p\partial

 \def\1{\bm1} 

\def\<{\left\langle}    \def\>{\right\rangle}
\def\({\left(}          \def\){\right)}
\def \[ {\left [} \def \] {\right ]}

\newcommand{\ve}{\varepsilon}

\newcommand{\B}[1]{{\bm{#1}}}%% Bold Roman & Greek Lower & Upper Case
\newcommand{\C}[1]{{\mathcal{#1}}}    %%   Calligrapfic Upper case
\renewcommand{\sb}[1]{_{\text {#1}}}  %% sub-   for lower case
\renewcommand{\sp}[1]{^{\text {#1}}}  %% super- for lower case
\def\Sb#1{_{\scriptscriptstyle\rm{#1}}}
\def\He4 {$^4$He~}

\begin{document}

\title{$^4$He Counterflow Differs Strongly from Classical Flows: Anisotropy on Small Scales}
\author{L. Biferale$^1$, D. Khomenko$^2$, V. L'vov$^3$, A. Pomyalov$^3$, I. Procaccia$^3$ and G. Sahoo$^4$}
\affiliation{$^1$Dept. of Physics, University of Rome, Tor Vergata,
	Roma, Italy\\
	$^2$ Laboratoire de physique th\'{e}orique, D\'{e}partement de physique de l'ENS, \'{E}cole normale sup\'{e}rieure, PSL Research University, Sorbonne Universit\'{e}s, CNRS, 75005 Paris, France.\\
	$^3$Dept. of Chemical and Biological Physics, Weizmann Institute of Science, Rehovot, Israel\\
	$^4$Dept. of Mathematics and Statistics and Dept. of Physics, University of Helsinki, Finland}
\begin{abstract}
	Three-dimensional anisotropic turbulence in classical fluids tends towards isotropy and homogeneity with decreasing scales, allowing --eventually-- the abstract model of ``isotropic homogeneous turbulence" to be relevant. We show here that the
	opposite is true for superfluid $^4$He turbulence in 3-dimensional counterflow channel geometry. This flow becomes less isotropic upon decreasing scales, becoming eventually quasi 2-dimensional. The physical reason for this unusual phenomenon is elucidated and supported
	by theory and simulations.
\end{abstract}

\maketitle
%\Fbox{comments by : \LB{Luca}, \AP{Anna}, \green{Victor}}
All turbulent flows in nature and in laboratory experiments are anisotropic on the energy injection scales \cite{2005-BP}. Nevertheless the model of ``isotropic homogeneous turbulence" had been shown to be highly
relevant and successful in predicting the statistical properties of turbulent flows on
scales much smaller than the energy injection scales (but still larger than the dissipative scales). The reason for this lies in the nature of the nonlinear terms of the equations
of fluid mechanics; these terms tend to isotropize the flow upon cascading energy to smaller scales, redistributing the anisotropic velocity fluctuations among smaller scales with a higher
degree of isotropy. Eventually, at small enough scales, the flow becomes sufficiently isotropic
to allow the application of the ideal model of isotropic homogeneous turbulence \cite{Frisch}.
In the present Letter,  we show that in turbulent superfluid $^4$He 
% the situation can be \red{different} in $^4$He superfluid turbulent flows. Here we find that 
in a channel geometry with a temperature gradient along the channel,  the opposite phenomenon takes place: the flow becomes less and less isotropic upon decreasing the scales. Eventually, the flow becomes quasi 2-dimensional with interesting and unusual properties as detailed below.

\begin{figure}
		\includegraphics[scale=0.25]{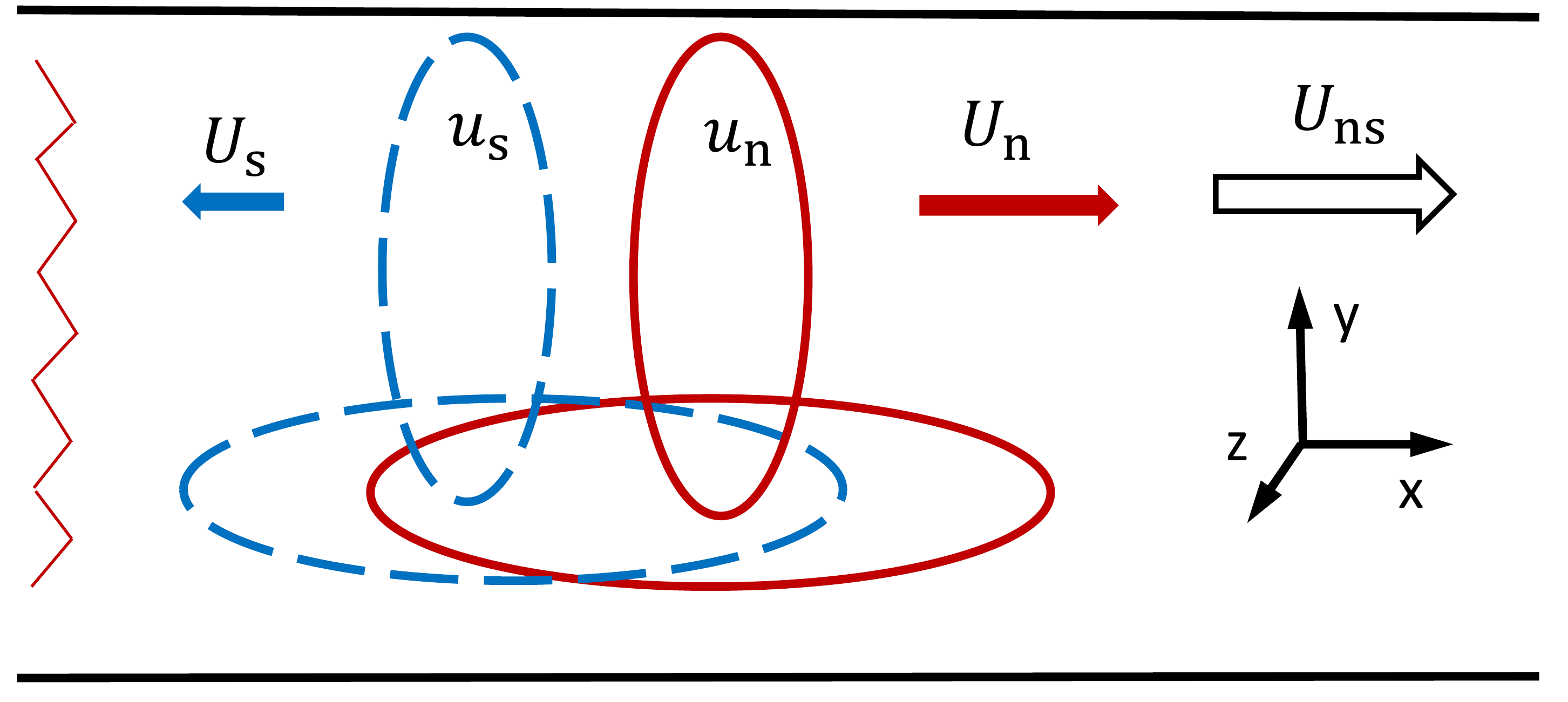}
	\caption{\label{f:diagram}
		Schematics  of the superfluid $^4$He channel counterflow. The normal-fluid eddies (solid red lines) and the superfluid eddies (the blue dashed lines) are swept by the corresponding mean velocities $U\sb n$ and $U\sb s$ away and towards  the heater, respectively. The resulting counterflow velocity $U\sb{ns}$ is oriented along the positive $x$-direction. The streamwise-elongated  eddies have longer overlap time than the cross-stream-elongated eddies. }  	
\end{figure}

An easy way to account for this difference in tendency towards isotropy is furnished
by the two-fluid model of turbulence in superfluid $^4$He \cite{Donnelly2009,Donnelly,Vinen}. Denote by $\B u_{\rm s}$ and $\B u_{\rm n}$ the superfluid and normal-fluid turbulent velocities, respectively. In counterflow geometry, with a temperature gradient directed along the channel, the mean superfluid velocity $\B U \sb s$ is directed towards the heater, and the mean normal velocity $\B U \sb n$ away from the heater. Importantly,  one finds that there exists a mutual
friction force $\B f_{\rm ns}$ between these two components \cite{Donnelly,2,Vinen,HV,Vinen3,37}, proportional to the difference in velocities, i.e
%\begin{equation}
$\B f_{\rm ns}\propto ~(\B u_{\rm n}-\B u_{\rm s} )$.
%\label{fns}
%\end{equation}
As long as the fluctuations between these two velocities are correlated, this force remains
small. Upon loss of correlation this force becomes large and will lead to a suppression
of the corresponding fluctuations. Consider then two types of velocity fluctuations, one
elongated along the channel and the counterflow and the other orthogonal to them, see
Fig.~\ref{f:diagram}. Due to the mean flow in opposite directions, the velocity
fluctuations oriented orthogonally will have a short overlap time and will decorrelate quickly, 
 whereas
the velocity fluctuations along the counterflow will remain correlated for a longer time.
The result will be a strong suppression of the former type of velocity fluctuations with
respect to the latter. This will eventually lead to a turbulent flow in which the fluctuations consist mostly of the stream-wise component, while the energy is concentrated in the plane orthogonal to the counterflow direction.
The rest of this Letter will elaborate
this  picture by using an analytical approach and will support it using direct numerical simulations (DNS).
%%%%%%%%%%%%%%%%%%%%%%%%%%%%%%%%%%%%%%%%%%%%%%%%%%%%%%%%%%%%%%
%%%%%%%%%%%%%%%%%%%%%%%%%%%%%%%%%%%%%%%%%%%%%%%%%%%%%%%%%%%%%%%%

{\sf The basic equations}.
The two-fluid model describes superfluid $^4$He of density $\rho$ as a mixture of two interpenetrating fluid components: an inviscid  superfluid and a viscous normal-fluid.  The densities of the components $\rho\sb s, \rho\sb n: \rho\sb s+ \rho\sb n=\rho$ define their contributions to the mixture. The fluid components  are coupled by a mutual friction force, mediated by the tangle of quantum vortices\,\cite{Donnelly,Vinen,HV,Vinen3,37} of a core radius $a_0\approx  10^{-8}\,$cm and  a fixed circulation $\kappa= h/M\approx 10^{-3}\,$cm$^2$/s, where $h$ is Planck's constant  and $M$ is the mass of the \He4 atom \cite{Feynman}.  A complex tangle of these vortex lines   with a typical inter-vortex distance \cite{Vinen}   $\ell\sim 10^{-4}- 10^{-2}\,$cm  is a manifestation of   superfluid turbulence.

To proceed it is sufficient to employ coarse-grained dynamics, following  the gradually-damped version\cite{He4} of the  Hall-Vinen-Bekarevich-Khalatnikov (HVBK)
equations for  counterflow turbulence \cite{He4,DNS-He4,LP-2018,DNS-He3,decoupling}.
It  has a form of two Navier-Stokes equations  for the turbulent velocity fluctuations $\B u_j(\B r,t)$  of the normal-fluid ($j=\rm n$) and the superfluid ($j=\rm s$):
 \begin{eqnarray}  \label{NSE} %%
 \Big[\frac{\p  }{\p t}+ (\B u_j +\B U_j)\*
\B\nabla\Big ] \B u_j
- \frac {\B \nabla p_j}{\rho_j }  =\nu_j\,  \Delta \B u_j  + \B f_j+ \B \varphi_j , %%
 \end{eqnarray} %%
coupled by the mutual friction forces $\B f_j$ in the
minimal form \cite{LNV}: $\B f\sb s \simeq  \Omega\sb s  \,(\B  u \sb n-\B  u \sb s )$,  $\B f\sb n \simeq   \Omega\sb n  \,(\B  u \sb s-\B  u \sb n )$, $\Omega\sb s =\alpha \kappa \C L$, and $\Omega\sb n  = \rho \sb s\Omega\sb s / \rho \sb n $. The mutual friction frequency $\Omega \sb s$ depends on  the
temperature-dependent dimensionless  mutual friction parameter $\alpha(T)$ and on the vortex line density $\C L$. In \Eqs{NSE} $p_j$ are the pressures of the normal-fluid and the
superfluid components. The kinematic
viscosity of the normal-fluid component is $\nu\sb n=\eta / \rho \sb n$ with $\eta$
being the dynamical viscosity of \He4 \cite{DB98}.  The  energy sink in the equation for the superfluid component, proportional to the effective superfluid viscosity, $\nu\sb s$,  accounts for the energy dissipation at the intervortex
scale $\ell$, due to vortex reconnections and energy transfer to Kelvin waves \cite{Vinen,He4}.   The contributions, involving  the  reactive
(dimensionless)   mutual friction  parameter $\alpha'$, that renormalizes  the
nonlinear terms, were omitted due to its numerical smallness \cite{DB98}. 

The large-scale motion in the thermal counterflow is sustained by the temperature gradient, created along the channel. Here we use the fact that the center of the channel flow at large enough Reynolds numbers  can be considered as almost space-homogeneous \cite{Pope}. To simplify the analysis we consider homogeneous turbulence  under periodic boundary conditions and  mimic the steering of turbulence at large scales by  random forces $\B \varphi_j$.
Equations \eqref{NSE} describe the  motion of two fluid components in the range of scales between the forcing scale and the intervortex distance.

{\sf Statistics   of anisotropic turbulence}.
The most general  description of  homogeneous superfluid \He4 turbulence at the level of second-order statistics can be done in terms of the three-dimensional (3D) Fourier-spectrum of each component  and the  cross-correlation functions:
\begin{equation} \label{eq:spectra}	
		(2\pi)^3\delta(\B k -\B k') \C F^{\alpha\beta}  _{ij} (\bm k)=\<  v _i^\alpha(\bm k)   v_j^{* \beta} (\bm k')\>, 		
\end{equation}
		 where  $\B v_j(\B k)$ is the Fourier transform  of  $\B u_j(\B r)$;  the indices $i$ and $j$ refer to the fluid components;  the vector indices $\alpha, \beta= \{x,y,z\}$ denote the  Cartesian coordinates and
  $^*$ stands for complex conjugation.  In the following,  we choose  the counterflow velocity, $U\sb{ns} = U\sb n-U\sb s$ along the $\hat {\B x}$-direction as depicted in  Fig.(\ref{f:diagram}).
  Next denote the trace of any tensor according to $\C F_{jj}(\bm k)\= \sum_{\alpha}  \C F^{\alpha\alpha}_ {jj}(\bm k)$. With this notation,
  the kinetic energy density per unit mass $\C E_j$ reads
\begin{equation}
    	\C E_j\= \frac12   \< |\B u _j(\B r)|^2 \>=   \frac 12  \int  \C F_{jj} (\bm k)  d^3 k \big /(2\pi)^3\, .
\end{equation}
Due to the presence of the preferred direction, defined by the counterflow velocity, 
the  counterflow turbulence has an axial symmetry around the $\hat {\B x}$ axis. Then  $\C F_{ij}(\B k)$   depends only on the two  projections $k_\|=k_x$ and $k_\perp=\sqrt{k_y^2+k_z^2}$  of the wave-vector $\B k$, being independent of the angle $\phi$ in the $\perp$-plane, orthogonal to $\B U\sb{ns}$.
This allows us to define a set of two-dimensional (2D) objects that still contain all the information about 2$\sp{nd}$-order statistics of the counterflow turbulence
\begin{subequations}\label{defs-F2}
\begin{equation}\label{def-F2A}
F_{ij} (k_\|,k_\perp) \=    \frac{k_\perp}{4\pi^2} \C F_{ij} (k_\|,k_\perp)\ .
	\end{equation}
	Another way to represent the same information is to introduce  a polar angle $cos(\theta)= ( \B k, \B U\sb{ns})/ |\B k|| \B U\sb{ns}| $,   and to use  spherical coordinates:
	\begin{eqnarray} \label{def-F2B}
 \tilde F_{ij} (k ,\theta) \=   \frac{k }{4\pi^2} \C F_{ij} (k\cos\theta,k\sin\theta) \  .  %\label{def-Fa}
	\end{eqnarray}	
\end{subequations}
{\sf Physical origin of the strong anisotropy}.
The physical origin of the strong anisotropy in the counterflow turbulence is best exposed by considering the  balance equation for the 2D energy spectra $\tilde F\sb{nn}(k,\theta), \tilde F\sb{ss}(k,\theta)$.
For that we start with \Eqs{NSE}, follow the procedure described in \,\Ref{LP-2018}
and average  the resulting equations for the 3D spectra   over the  azimuthal  angle $\varphi$.
Finally, for the normal component we get:
 	\begin{eqnarray}\label{balance1} \nonumber
	 &&\frac{\partial  \tilde F_{\rn \rn}(k,\theta,t)}{\partial t} \!+\!	\mbox{div}_{\B k } [\B \ve_\rn ( \B k)] = - \C D_\rn\sp{mf}(k,\theta ) \!-\! \C D_\rn\sp{kv}(k,\theta ),\\
 &&	\C D_\rn\sp{mf}(k,\theta )=\Omega_\rn [ \,\tilde F_{\rn \rn}  (k,\theta)- \tilde F\sb{ns}(k,\theta)  \big ]\, , \\ \nonumber
 &&\C D_\rn\sp{kv}(k,\theta )= 2\, \nu_\rn k^2 \,\tilde F_{\rn\rn }(k,\theta)\,,
 	 	\end{eqnarray}
        where div$_{\B k} [\B \ve_j ( \B k)]$ is the transfer term due to inertial non-linear effects,  $\C D_\rn\sp{mf}(k,\theta )$
        %=\Omega_\rn [ \,\tilde F_{\rn \rn}  (k,\theta)- \tilde F\sb{ns}(k,\theta)  \big ] $
         describes the rate of energy dissipation  by the mutual friction, while $\C D_\rn\sp{kv}(k,\theta )$
         %= 2\, \nu_\rn k^2 \,\tilde F_{\rn\rn }(k,\theta)$ 
         stands for the rate of dissipation  by  the kinematic viscosity. A similar equation is obtained  for the superfluid component by replacing $\rn$ with $\rs$ everywhere.
%%%%%%%%%%%%%%%%%%%%%%%%%%%%%%%%%%%%%%%%%%%%%%%%%%%%%%%%%%%%%%%%
\begin{figure*}
	\begin{tabular}{ccc}
			\includegraphics[scale=0.27]{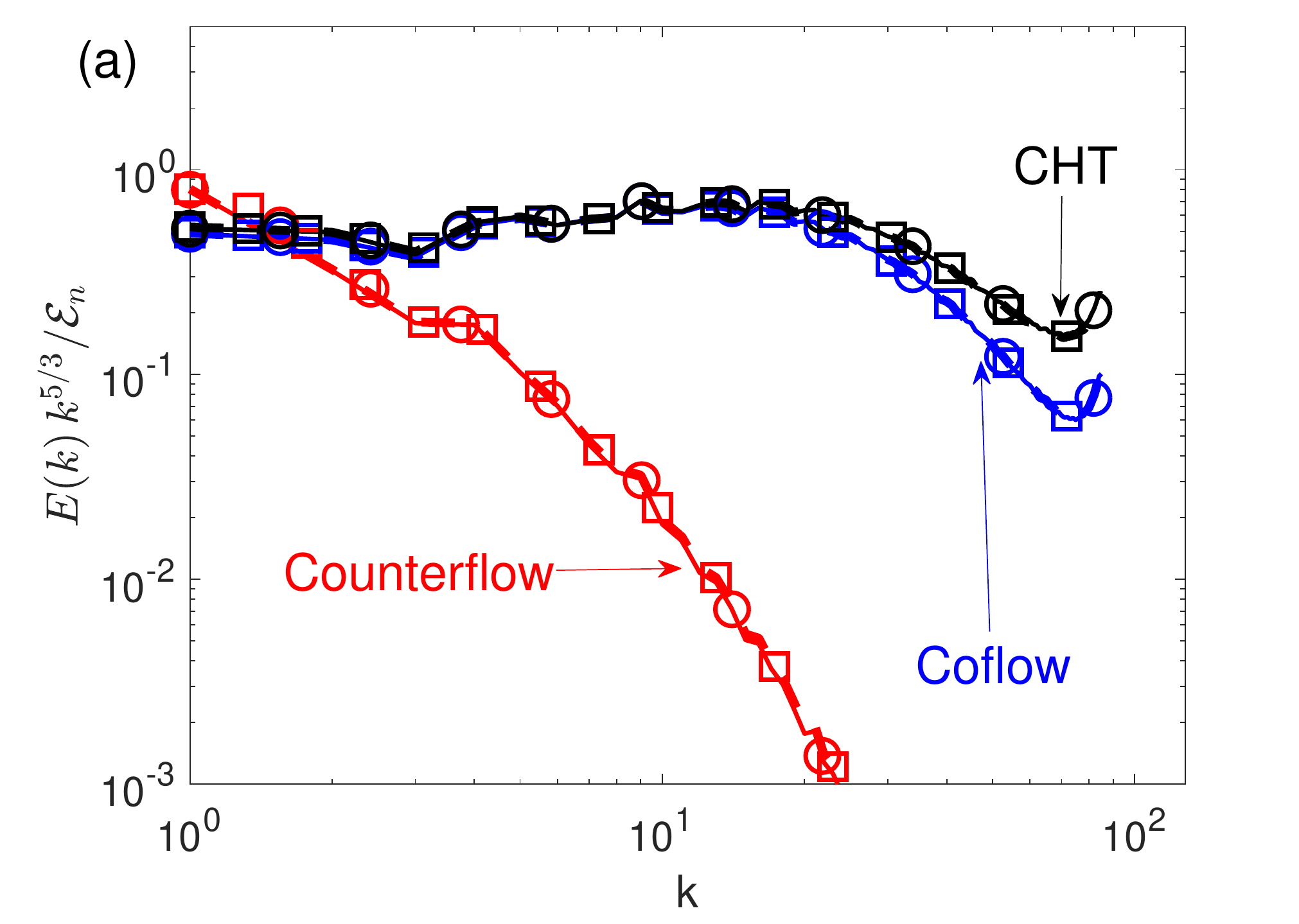}&
		\includegraphics[scale=0.27]{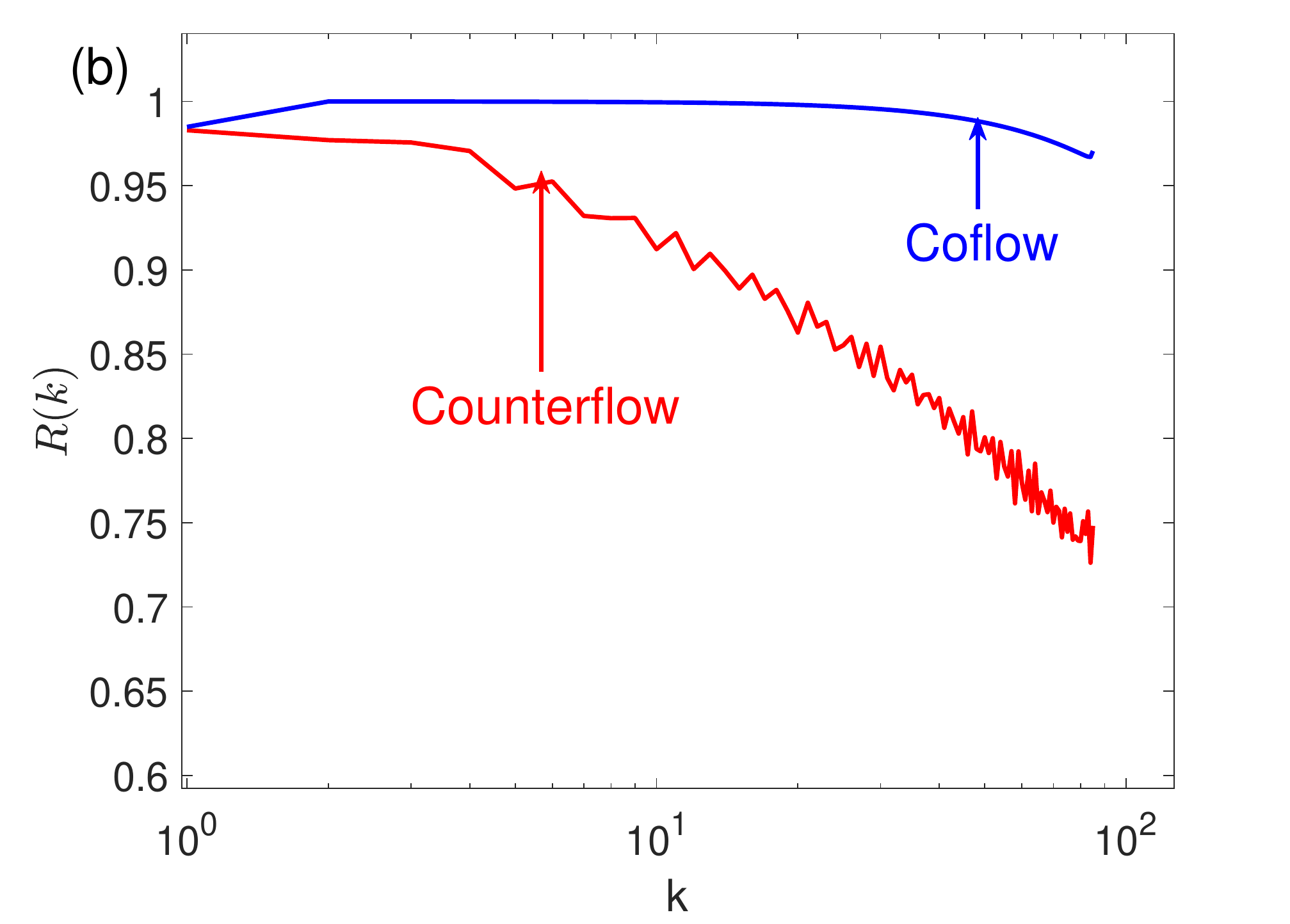}&
		\includegraphics[scale=0.27]{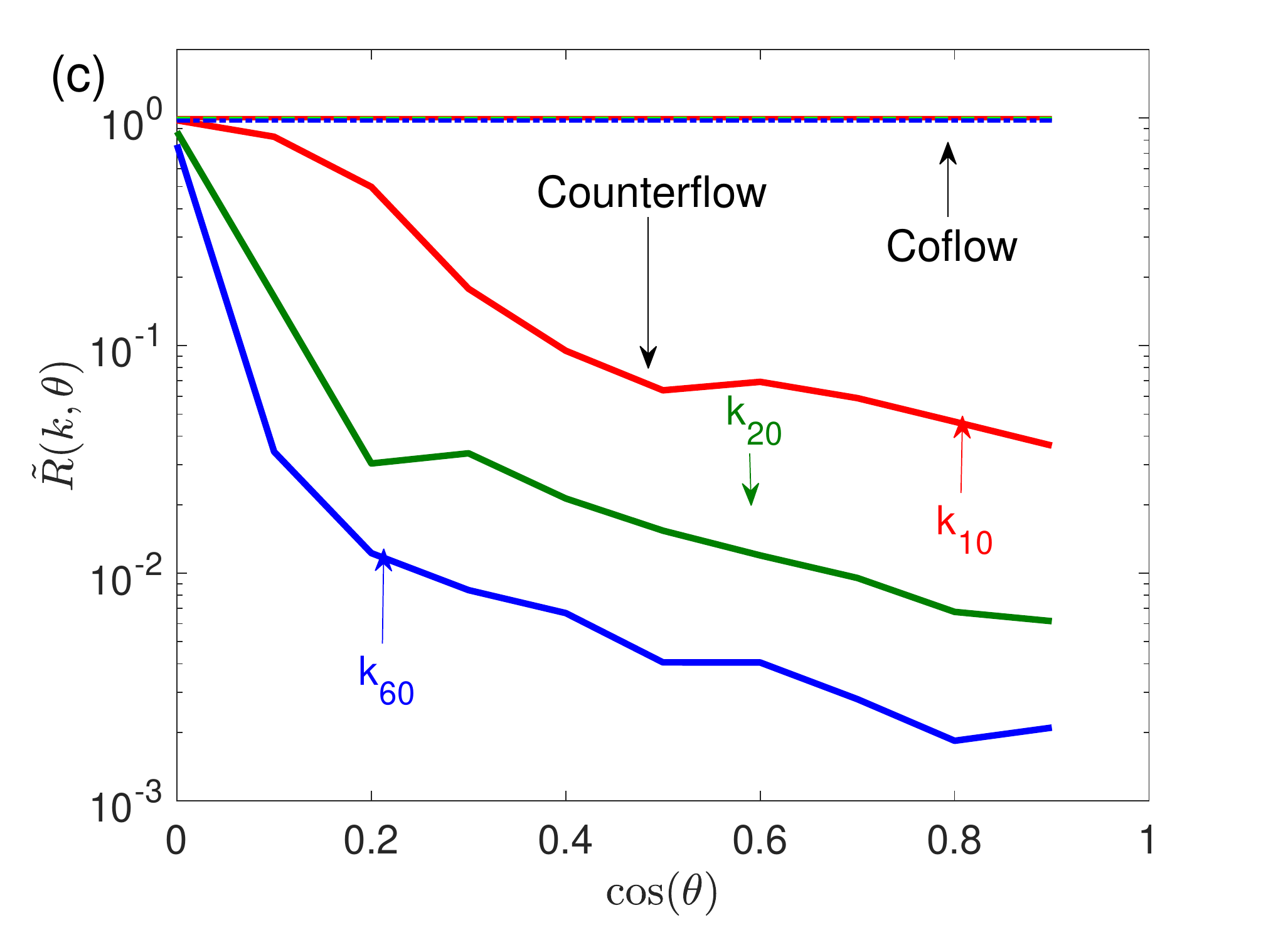}
	\end{tabular}
	\caption{ \label{f:1}(a) The spherical energy spectra $E_{jj}(k)$ of the normal-fluid (circles) and the superfluid (squares), (b) the cross-correlation function  $R(k)$ and (c)  the angular dependence of the  cross correlation function $\tilde R(k,\theta)$  for the coflow  and the counterflow. In panel (c), the data for the coflow all coincide with the isotropic result. For the counterflow, red lines correspond to the $\tilde R(k,\theta)$  averaged over the wavenumber range $10\leq k< 20$, green lines -- to averaging over $20\leq k< 60$  and blue lines -- to the averaging over $60\leq k\le 80$  (labeled as $k_{10}$,  $k_{20}$, and $k_{60}$, respectively).   Note the log-linear scale.  }
\end{figure*}
%%%%%%%%%%%%%%%%%%%%%%%%%%%%%%%%%%%%%%%%%%%%%%%%%%%%%%%%%%%%%%%%
For a qualitative analysis of the origin of the anisotropy in our system it is important to develop a closure of the cross-correlation function $\tilde F\sb{ns}(k,\theta)$ in   $\C D_j\sp{mf}(k,\theta )$ in terms of the spectral properties of each fluid component and of the counterflow velocity.

 According to \Ref{decoupling}:
	 \begin{equation}\label{LP-20A}
\tilde F\sb{ns}(k,\theta)=  AB/ [B^2+ (\B k\cdot \B U\sb{ns})^2]\ .
	\end{equation}
	Here $A= \Omega\sb s \tilde F\sb{nn} (k,\theta)+ \Omega\sb n \tilde F\sb {ss} (k,\theta)$
 and $B$ 	can be approximated  as $B=\Omega\sb n + \Omega \sb s$, as shown  in \cite{LP-2018}.
We further simplify $\tilde F\sb{ns}(k,\theta)$ in \Eqs{LP-20A} by noting \cite{LP-2018} that when two components are highly correlated, the cross-correlation may be accurately represented by the corresponding energy spectra. For  wavenumbers where the components are not correlated, as is quantified by the decorrelation function $D(k,\theta)$ \cite{decoupling}, $\tilde F\sb{ns}(k,\theta)$  is small and the accuracy of its representation is less important. We therefore get a decoupled form of the cross-correlation:
	\begin{subequations}\label{LP-20}
	\begin{eqnarray}\label{LP-20B}
\tilde F\sb{ns}(k,\theta)&=&\tilde F_{j j}(k,\theta) D(k,\theta)\,,  \\ \label{LP-20C}
	D(k,\theta) &=& \Big [1+ \Big (\frac{k U\sb {ns}\cos \theta}{\Omega\sb{n}+\Omega\sb{s}}\Big )^2 \Big]^{-1} \,,
	\end{eqnarray}
	 and finally determine the rate of energy dissipation due to mutual friction:	
\begin{equation}\label{diss}
\C D_j\sp{mf}(k,\theta)= \Omega_j \tilde F_{j j}(k,\theta) \big [ 1- D(k,\theta) \big]\ .
\end{equation}
\end{subequations}
 Equations \eqref{LP-20} are the central analytical result of this paper.

 The impact of $U\sb{ns}$ on the anisotropy  follows from the closure \eqref{diss}. Indeed, for small $k$ or even for large $k$ with $\B k$  almost perpendicular to  $\B U\sb{ns}$ (i.e $\cos\theta\ll 1$), $D(k,\theta)\simeq 1$, the normal-fluid and superfluid velocities are almost fully coupled and the dissipation rate  is small: $\C D_j\sp{mf}(k,\theta)\ll \Omega_j$.   In this case, the mutual friction does not significantly affect the  energy balance and we  expect the energy spectrum  $\tilde F_{jj}(k,\theta)$ to be close to the Kolmogorov-1941 (K41) prediction
$E\Sb{K41}(k)\propto k^{-5/3}$ for both components.   For large $k$ and with $\cos\theta\sim 1$, the  velocity components are almost decoupled $D(k,\theta) \ll  1$, and the mutual-friction energy dissipation is maximal: $\C D_j\sp{mf}(k,\theta)\approx \Omega_j\~ F_{jj}(k,\theta)$. This situation is similar  to that in $^3$He with the normal-fluid component at rest \cite{DNS-He3}.   In such a  case, we can expect that the energy dissipation by mutual friction  strongly suppresses the energy spectra, much below the K41 expectation $E\Sb{K41}(k)$.
%%%%%%%%%%%%%%%%%%%%%%%%%%%%%%%%%%%%%%%%%%%%%%%%%%%%%%%%%%%%%%%%%%%%%%%%%%%%
\begin{figure*}
%\hskip -0.7cm \includegraphics[scale=0.36]{E_2d_kpkx_Fig3a.eps}
%	\hskip-0.7cm		\includegraphics[scale=0.29]{Exyz_Fig3b.eps}	
%	\hskip 0.5cm 	\includegraphics[scale=0.48]{Fig3c.eps}	
	\hskip -0.7cm \includegraphics[scale=0.36]{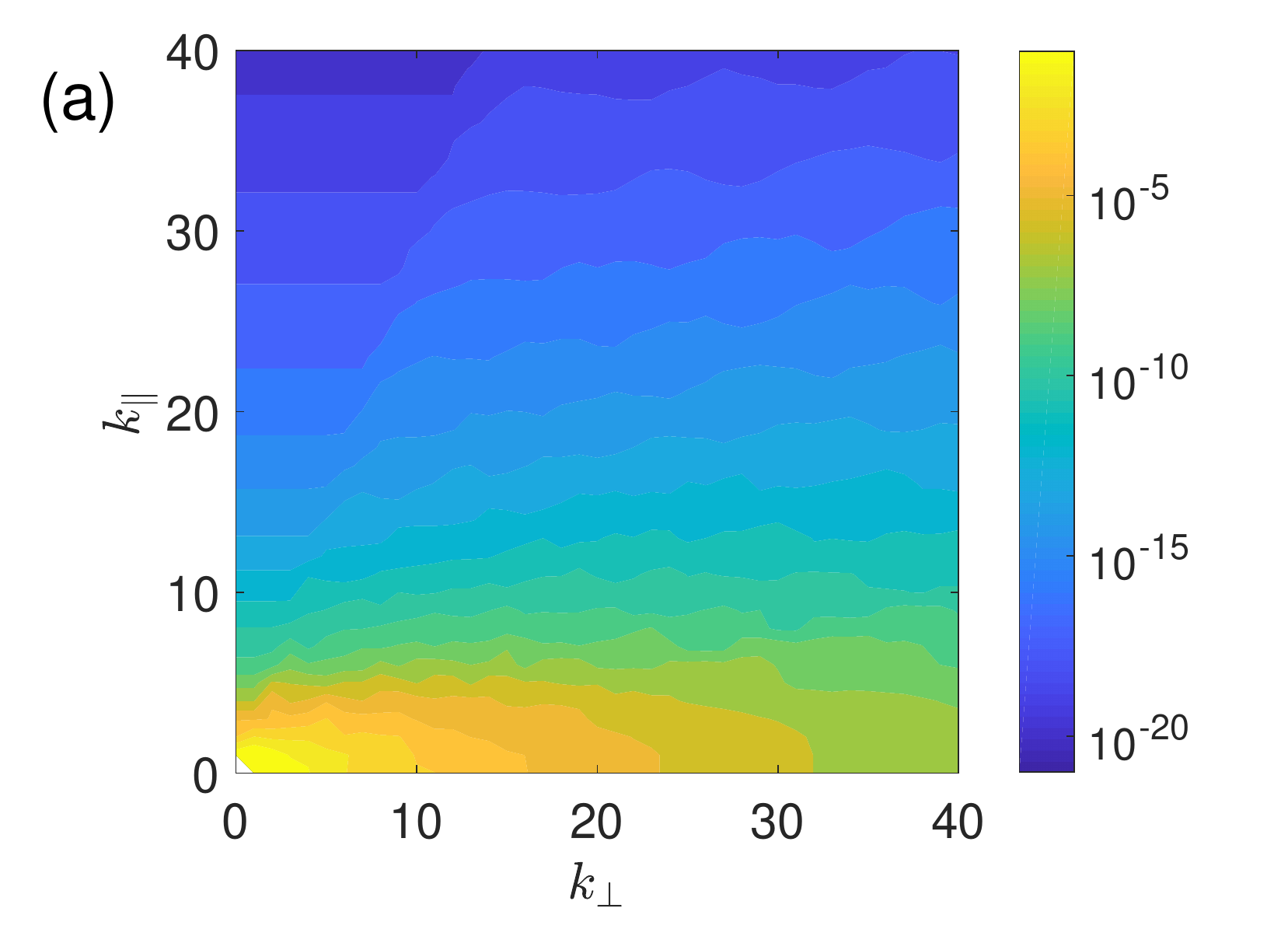}
	\hskip-0.7cm		\includegraphics[scale=0.29]{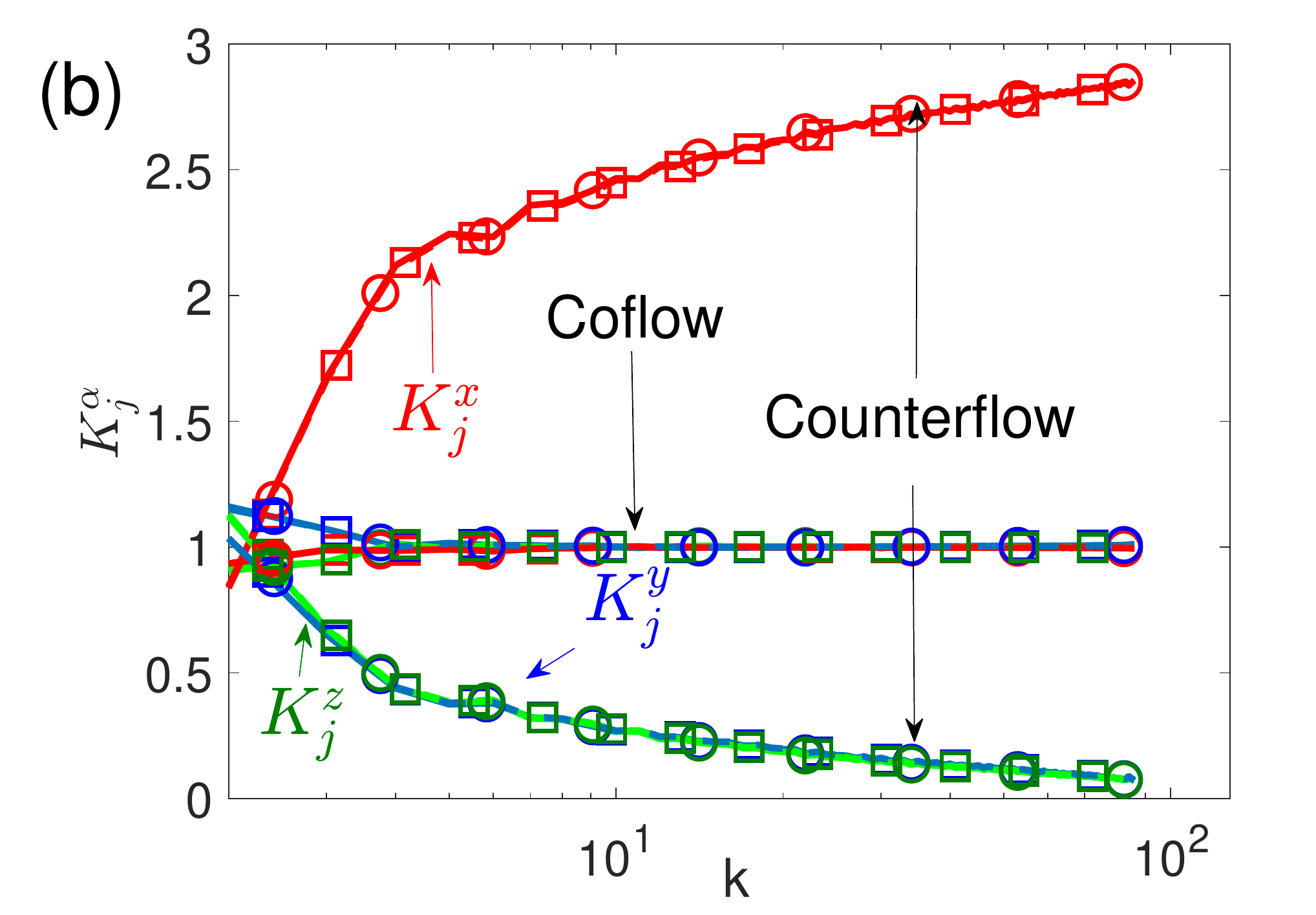}	
	\hskip 0.5cm 	\includegraphics[scale=0.48]{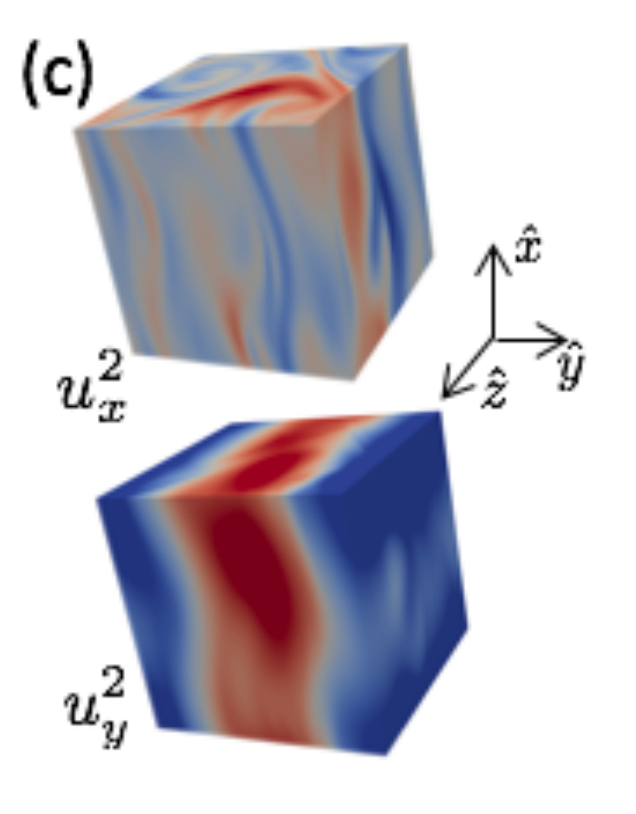}	
	\caption{\label{f:2}
		(a)The  superfluid component energy spectrum $ F\sb {ss} (k_\|,k_\perp)$ in the counterflow. (b) The tensor decomposition of the normalized spherical  energy spectra $K_j^\alpha(k)$ for the normal-fluid  (circles) and the superfluid (squares). (c) The superfluid velocity components $\B u_{\rm  s}^x(\B r)$(top) and  $\B u_{\rm  s}^y(\B r)$(bottom). The $\B u_{\rm  s}^z(\B r)$ (not shown) is similar to $\B u_{\rm  s}^y(\B r)$. The velocity magnitude is color-coded with red denoting positive and blue denoting negative values.
	}  	
\end{figure*}
%%%%%%%%%%%%%%%%%%%%%%%%%%%%%%%%%%%%%%%%%%%%%%%%%%%%%%%%%%%%%%%%%%%%%%%%%%%%%%%%%%%%%%%%%
Combining all these considerations, we expect the energy spectra $\tilde F_{jj}(k,\cos \theta)$  to become more  anisotropic with increasing $k$, with most of the energy concentrated in the range of small $\cos \theta$, i.e. in the orthogonal plane.

{\sf Numerical results}.
Direct numerical simulations of the coupled HVBK \Eqs{NSE} were carried out using a
fully de-aliased pseudospectral code with a resolution of $256^3$ collocation points in a triply
periodic domain of size $L=2\pi$.
To reach a steady state flow, velocity fields of the normal and superfluid components are stirred by two independent random Gaussian forces $\B \varphi\sb s$ and $\B \varphi\sb n$ with the force amplitudes $|\varphi|=0.5$  for both components, localized  in the band  $k_{\varphi}\in [0.5,1.5]$. The time integration is performed using  2-nd order Adams-Bashforth scheme with viscous term exactly integrated.

We have decided to focus on the temperature $T=1.85$\,K, at which the densities and viscosities of the normal-fluid and superfluid components are close: $\rho\sb s/\rho\sb n=1.75$ and $\nu\sb s/\nu\sb n=1.07$. The mutual friction parameter for this temperature is $\alpha=0.18$.  The simulations  were carried out with both the  normal-fluid and superfluid viscosity  $\nu\sb n=\nu\sb s=0.003$.
%and the value of  $\nu\sb s=0.00185$  
%was found using the known value of ratio $\nu\sb s/\nu\sb n$.
	Other  parameters of the simulations were chosen based on the relevant dimensionless relations: the Reynolds numbers 
	%$Re_j=(u\sp j \Sb T)(/\nu_j k_0$  
	and the normal-fluid turbulent intensity $w$
	\begin{equation}
	    Re_j=(u^j \Sb T)(/\nu_j k_0)\,,\quad  w=U\sb{ns}/ u\sp n \Sb T\, .
	\end{equation}
%		and  the turbulent intensity $U\sb{ns}/ u\sp n \Sb T$.
Here  $u^j \Sb T=\sqrt {\langle {u_j}^2 \rangle} $ is the root mean square (rms)  of the turbulent velocity fluctuations, $k_0=1$  is the outer scale of turbulence.	
 To emphasize the importance of the counterflow, we compare the results with the  simulations for the so-called coflow  with the rest of the parameters being  the same. In the coflow, the two components of the mechanically driven   $^4$He, being coupled by the  mutual friction force,  move in the same direction with the same mean velocities, $U\sb{ns}=0$. The statistics in the coflow configuration is known to be similar to that of  classical isotropic turbulence \cite{DNS-He4,TenChapters,BLR,Roche-new}.
 In our simulations, the values of the Reynolds numbers in the counterflow are  $Re\sb n=1051 $ and $Re\sb s=1056$, while in the coflow, $Re\sb n=1179 $ and $Re\sb s=1181 $. The rms velocities of both components in both flows are   $u\sp s \Sb T=  u\sp n \Sb T  =3.5$.
 The dimensionless values of the mutual friction frequency $\Omega\sb s=20$ and the counterflow velocity $U\sb{ns}=15.4$  correspond to  the case with both components strongly turbulent and strongly coupled.
The results on the temperature and $\Omega\sb s$ dependence of the energy spectra will be reported elsewhere.
The flow conditions were controlled by the simulations of the  uncoupled equations without counterflow ($\B U_j = \Omega_j=0$), which represent here the classical hydrodynamic isotropic turbulence (CHT).

The energy spectra are influenced by the viscous dissipation, by the  dissipation due to mutual friction and by the counterflow-induced decoupling.
To clarify the role of each of these factors, we first ignore the expected anisotropy and compare  in \Fig{f:1}(a) the  normal-fluid and superfluid   energy spectra $E\sb {nn}(k)$ and $E\sb {ss}(k)$ and the cross-correlation $E\sb {ns}(k)$, integrated over a spherical surface of radius $k$, i.e. over all directions of vector $\B k$:
\begin{eqnarray} \label{Esp}
 E_{ij} (k)=  \int \C F_{ij}(\B k) \, \frac{d\phi\,  d\cos \theta }{(2\pi)^3}\ .
\end{eqnarray}
  The corresponding  normalized cross-correlation functions
 \begin{equation}
     R(k)=2\, E\sb{ns}(k)/[E\sb {nn}(k)+  \, E\sb {ss}(k)]
 \end{equation} 
   are shown in \Fig{f:1}(b).
%%%%%%%%%%%%%%%%%%%%%%%%%%%https://www.overleaf.com/project/5bfe92b5c924f57345ad7b98
	 The effect of viscous dissipation is clearly seen in the spectra of the uncoupled components, corresponding to  classical hydrodynamic turbulence (marked ``CHT", black lines). The spectra almost coincide, since at $T=1.85$\,K the viscosities are close.
	In the coflow, the strongly coupled components are well correlated at all scales and move almost as one fluid.  Note the additional dissipation due to mutual friction, leading to further suppression of the spectra compared to the uncoupled case.
	 The presence of the counterflow velocity leads to a sweeping \cite{decoupling} of the two component's eddies in opposite directions by the corresponding mean velocities. The result is the decorrelation of the components turbulence velocities, especially at small scales, for which the overlapping time is very short, see \Fig{f:1}(b).  The dissipation by mutual friction is very strong in this case, with both $\Omega$ and the velocity difference being large, leading to very strongly suppressed spectra, with $ E\sb{nn} (k)\approx E\sb {ss}(k)$. This behavior was predicted by the theory \cite{LP-2018}, based on the assumption of spectral isotropy.
However the spherically integrated spectra and cross-correlations cannot reveal any properties connected to the anisotropic action of the mutual friction force.  To account for the spectral anisotropy we plot  in \Fig{f:1}(c) the normalized 2D cross-correlations
	\begin{equation}
\tilde R(k,\theta)=2\tilde F\sb{ns}(k,\theta) /[\tilde F\sb{nn}(k,\theta)+  \tilde F\sb{ss}(k,\theta)].
	\end{equation}
Given the discrete nature of the $\B k$-space in DNS, we average them over 3 bands of wavenumbers. Leaving aside  $k\approx k_0$, influenced by the forcing, we average $\tilde R(k,\theta)$ over the $k$-ranges $10\leq k< 20$,  $20\leq k< 60$  and   $60\le k \le 80$.

The first observation here is that the cross-correlation for the coflow are
	isotropic at all scales, see thin horizontal lines, marked ``coflow". On the other hand, in the counterflow, the cross-correlations are largest for $\cos\theta \approx 0$ and fall off very fast with decreasing angle, slower for small $k$ (red lines, labelled $k_{10}$) and faster as $k$ become larger (green, $k_{20}$, and blue lines, $k_{60}$, respectively). Such a strong decorrelation of the components velocities leads to an enhanced dissipation by mutual friction in the counterflow direction, such that most of the energy
	is contained in the narrow range $\cos\theta \lesssim0.1$, near the
	plane orthogonal to $\B U\sb{ns}$.

Indeed, the superfluid energy spectrum $F\sb{ss} (k_\|,k_\perp)$, shown in \Fig{f:2}(a),  is strongly suppressed in the $k_{||}$ direction, while it decays slowly in the orthogonal plane. A similar  phenomenon of the creation of quasi-2D turbulence is observed  in a strongly stratified atmosphere \cite{AtmTurb-review,2018-AB,atmosTurb-Kumar} and in rotating turbulence\,\cite{rot1,rot2,rot3}, in which there exists a preferred  direction defined by gravity or by a rotation axis.  The difference between these examples  and the present counterflow lies in the nature of the velocity field. The leading velocity components in the classical flows are in a plane orthogonal to the preferred direction. Moreover, at small scales the isotropy is  restored \cite{atmosTurb-Kumar,2018-AB}. On the contrary, in $^4$He counterflow, the dominant velocity component is oriented along the counterflow direction, with the anisotropy becoming stronger with decreasing scales, as we show in  \Fig{f:2}(b). Here we plot the tensor components of the spherical spectra as the ratios
	\begin{equation}\label{tensor_ratio}
	K_j^\alpha(k) \= 3\, E^{\alpha\alpha}_{jj}(k)/\, E_{jj}(k).
	\end{equation}
	The factor 3 was introduced to ensure that for  isotropic turbulence   $K_j^\alpha(k)=1$.
	Expectedly,  the coflow (the almost horizontal lines) is isotropic at all scales, except for the smallest wavenumbers.
	On the other hand, for the counterflow turbulence, the contribution of the  $K_j^x(k)$ component (shown by red lines) is dominant and monotonically increases with $k$ from the isotropic level $K_j^x(k_0)\approx 1$ to the maximal possible level $K_j^x(k )\approx 3$. Therefore the small-scale counterflow turbulence consists mainly  of $v_j^x(k)$ velocity fluctuations. The  contribution of $v_j^y$ and $v_j^z$ fluctuations for $k\gtrsim 10$ is negligible.
	Summarizing \Fig{f:2}, the leading contribution to the spectra of small scale counterflow turbulence comes from the turbulent  velocity fluctuations with only one stream-wise projection that depends on the two cross-stream coordinates $\{y,z\}$: $ u^x(y,z)$.
 Such type of turbulence can be visualized as narrow jets or thin sheets with velocity, oriented along the counterflow and  randomly distributed in the $\perp$-plane. Indeed
	  the velocity components $u^y_{\rm s}$, shown in \Fig{f:2}c, and $u^z_{\rm s}$ have only large scale structures, while $u^x_{\rm s}$ has elongated structures at various scales.  The energy spectra, corresponding to $ u^x_{\rm n}(y, t)$ were recently measured experimentally \cite{WG-2017,WG-2018} and were found to agree with predictions \cite{LP-2018} in the range of scales where the fluid components are well correlated, while decaying faster than predicted at smaller scales.
	  
{\sf Summary}.
The  energy spectra  of the  superfluid $^4$He counterflow turbulence  become  more anisotropic  upon going from large scales toward scales about the intervortex distance. This strong anisotropy distinguish it from the classical turbulent flows that become more isotropic as the scale decreases.  Most of the turbulent energy become concentrated in the plane, orthogonal  to the counterflow direction. Furthermore, contrary to  classical  quasi-2D turbulent flows  in rotation or in stratified configurations, where  dominant velocity components lie in the same plane, the only surviving velocity component at small scales  is preferentially oriented along the counterflow direction.
The selective suppression of the orthogonal velocity fluctuations has its origin in the strong anisotropy of the energy dissipation by mutual friction, resulting from the angular dependence of the components' cross-correlation. 

\textbf{Acknowledgments}
LB acknowledges funding from the European Unions Seventh Framework Programme (FP7/20072013) under Grant Agreement No. 339032. GS thanks AtMath collaboration at University of Helsinki. DK acknowledges funding from the Simons Foundation under grant No. 454955 (Francesco Zamponi).


\begin{thebibliography}{99}
		\bibitem{2005-BP} L. Biferale and I. Procaccia, 
		%Anisotropy in Turbulent Flows and in Turbulent Transport,
		Phys. Rep. \textbf{414} 43, (2005).
	\bibitem{Frisch} U. Frisch, Turbulence, the legacy of A.N. Kolomogorov, Cambridge Univ. Press, 1995.
	\bibitem{Donnelly2009}R. J. Donnelly, 
	%The two-fluid theory and second sound in liquid helium, 
	Physics Today \textbf{62},  34 (2009).
	\bibitem{Donnelly}R. J. Donnelly, \textit{ Quantized Vortices in Hellium II }(Cambridge 3 University Press, Cambridge, 1991).

	\bibitem{Vinen} W. F. Vinen and J. J. Niemela,  
	%Quantum turbulence.
	J. Low Temp. Phys. \textbf{128}, 167 (2002).
	\bibitem{HV} H. E. Hall and W. F. Vinen, 
	%The rotation of liquid helium II. I. Experiments on the propagation of second sound 	in uniformly rotating helium II.
	Proc. Roy. Soc. A \textbf{238}, 204 (1956).
	
	\bibitem{Vinen3} W. F. Vinen,
	%Mutual friction in a heat current in liquid helium 	II I. Experiments on steady heat currents, 
	Proc. R. Soc. \textbf{240},	114 (1957); 
	%Mutual friction in a heat current in liquid helium II.	II. Experiments on transient effects, 
	\textbf{240}, 128 (1957); 
	%Mutual friction in a heat current in liquid helium II III. Theory ofthe mutual friction, 
	\textbf{242}, 493 (1957); 
	%Mutual 	friction in a heat current in liquid helium. II. IV. Critical heat	currents in wide channels,
	\textbf{243}, 400 (1958).
	
	\bibitem{37} R. N. Hills and P. H. Roberts, 
	%Superfluid mechanics for a high	density of vortex lines, 
	Arch. Ration. Mech. Anal. \textbf{66}, 43 (1977).
	
	\bibitem{2} \emph{Quantized Vortex Dynamics and Superfluid Turbulence}, edited by C.F. Barenghi, R.J. Donnelly and W.F. Vinen, Lecture Notes in Physics	\textbf{571} (Springer-Verlag, Berlin, 2001)
	
	\bibitem{Feynman} R. P.Feynman,
	%Application of quantum mechanics to liquid helium.
	Progress in Low Temperature Physics \textbf{1}, 17 (1955).
	
	\bibitem{He4} L. Boue, V.S. L'vov, Y. Nagar, S.V. Nazarenko, A. Pomyalov, I. Procaccia,  
	%Energy and vorticity spectra in turbulent superfluid He-4 from $T=0$ to $T_\lambda$.
	Phys. Rev. B. \textbf{91},  144501, (2015).
		
	\bibitem{decoupling} D. Khomenko, V. S. L'vov, A. Pomyalov, and I. Procaccia,
	%Counterflow induced decoupling in superfluid Turbulence.  
	Phys. Rev. B \textbf{93},  014516 (2016).
				
	\bibitem{DNS-He3} L. Biferale, D. Khomenko, V. L'vov, A. Pomyalov, I. Procaccia and G. Sahoo, 
	%Local and non-local energy spectra of superfuid $^3$He turbulence,
	Phys. Rev. B. \textbf{95}, 184510 (2017).
		
	\bibitem{DNS-He4} L. Biferale,  D. Khomenko,  V.S.  L'vov,  A. Pomyalov,  I. Procaccia,  and G. Sahoo,	
	%Turbulent statistics and intermittency enhancement in coflowing superfluid	$^4$He, 
	Phys. Rev.Fluids \textbf{3}, 024605 (2018).
		
	\bibitem{LP-2018} V. S. L'vov and  A. Pomyalov, 
	%A theory of counterflow velocity dependence of superfluid $^4$He turbulence statistics,
	Phys. Rev. B, \textbf{97}, 214513 (2018).
	\bibitem{LNV} V. S. L'vov, S. V. Nazarenko and G. E. Volovik, 
	%Energy spectra of developed superfluid turbulence,
	JETP Letters, \textbf{80},  535 (2004).
	\bibitem{DB98}  R. J. Donnelly, C. F. Barenghi , 
	%The Observed Properties of Liquid Helium at the Saturated Vapor Pressure,
	J. Phys. Chem. Ref. Data \textbf{27,} 1217(1998).
	\bibitem{Pope}S. B. Pope, \textit{Turbulent Flows} (Cambridge University Press,	Cambridge, 2000).	
	\bibitem{TenChapters} L. Skrbek and K. R. Sreenivasan, in \emph{Ten Chapters in Turbulence},	edited by P. A. Davidson, Y. Kaneda, and K. R. Sreenivasan	(Cambridge University Press, Cambridge, 2013), pp. 405--437.
				
	\bibitem{BLR} C. F. Barenghi, V. S. L'vov, and P.-E. Roche, %Experimental, numerical, and analytical velocity spectra in turbulent quantum fluid,
	Proc Natl Acad Sci USA \textbf{111}, 4683  (2014).
	
	\bibitem{Roche-new}E. Rusaouen, B. Chabaud, J. Salort, Philippe-E. Roche.
	%Intermittency of quantum turbulence with superfluid fractions from 0\% to 96\%.
	Physics of Fluids  \textbf{29}, 105108 (2017).
							
	\bibitem{AtmTurb-review}	E.J. Hopfinger,
	%Turbulence in stratified fluids: a review.
	J Geophys Res. \textbf{92},5287(1987).
					
	\bibitem{atmosTurb-Kumar}	A. Kumar, M. K. Verma and J. Sukhatme,
	%Phenomenology of two-dimensional stably stratified turbulence under large-scale forcing, 
	J. of Turbulence, \textbf{18}, 219(2017).
	
	\bibitem{2018-AB}		A. Alexakis, L. Biferale, 
	%Cascades and transitions in turbulent flows, 
	Phys. Rep. \textbf{767}-\textbf{769},1 (2018).
				
	\bibitem{rot1}	 L. Biferale, F. Bonaccorso, I. M. Mazzitelli, M. A. T. van Hinsberg, A. S. Lanotte, S. Musacchio, P. Perlekar, and F. Toschi. 
	%Coherent Structures and Extreme Events in Rotating Multiphase Turbulent Flows. 
	Phys. Rev. X \textbf{6}, 041036 (2016).
	
	\bibitem{rot2} B. Gallet, A. Campagne, P.-P. Cortet, and F. Moisy, %Scale-Dependent Cyclone-Anticyclone Asymmetry in a Forced Rotating Turbulence Experiment, 
	Phys. Fluids \textbf{26}, 035108 (2014).
	
	\bibitem{rot3}. B. Gallet,
	%Exact Two-Dimensionalization of Rapidly Rotating Large-Reynolds-Number Flows,
	J. Fluid Mech. \textbf{783}, 412 (2015).
				￼
	\bibitem{WG-2017} J. Gao,  E. Varga, W. Guo and W. F. Vinen,
	%Energy spectrum of thermal counterflow turbulence in superfluid Helium-4,
	Phys. Rev. B \textbf{96}, 094511 (2017).
	
	\bibitem{WG-2018}S. Bao, W. Guo,  V. S. L'vov, A. Pomyalov, %Statistics of turbulence and intermittency enhancement in  superfluid $^4$He counterflow , 
	Phys. Rev. B   \textbf{98}, 174509 (2018).
		
\end{thebibliography}
\end{document}